# Strategies for Development of a Distributed Framework for Computational Sciences[*].

Vladimir Berezovsky, Alexander Popov,

Pomor State University named after M.V.Lomonosov,
Department of Mathematics,
Uritskiy str., 69 B,
163002 Arkhangelsk, Russia
vladimir@ltu.se, aleneus@gmail.com

**Abstract.** This paper discusses some generic approach for developing grid-based framework for enabling establishment of workflows comprising existing software in computational sciences areas. We highlight the main requirements addressed the developing of such framework. Some strategies for enabling interoperability between convenient computation software in the grid environment has been shown. The UML based instruments of graphical description of workflows for the developing system has been suggested.

## 1  Introduction

Nowadays, the computational physic, chemistry and material sciences has an important place in the theoretical sciences, building a bridge to the experimental science, allowing to solve the complex problems that can't be solved analytically. Mainly, the scientific computing concerned with construction of mathematical models and using computers to do the quantitative analysis of this model during the computer simulation. At this time, there are a lot of computational software dedicated to solving a particular science problem using one or another methodology. Many complex computational problems can be decomposed into the multiple subproblems, and this demands to organize the interaction of particular computational applications one with other to solve the target problem. This leads to the formation of the computational workflow that consists of the application chain, where results from the output of particular application put onto the input of the other one next in the chain. Such problems as multiscale modeling, complex computational simulations, fitting parameters from experimental data, comparing and evaluation of computational methodologies, verification of numerical solutions encourages to formation of these chains. For example, let us consider the helium permeation through the zeolite pores in the presence of a heavy adsorbing compound. One of the methods for modeling of such system can be done in following steps. The Crystallographic data for zeolite lattice can be taken from the Database of zeolite structures[1]. Then the displacement of heavy adsorbing compound molecules in the zeolite can be obtained using configurational-bias Monte Carlo (CBMC) approach. One can use the BIGMAC code[2] for performing CBMC simulations in grand canonical ensemble for this. Next step is distribution of the helium atoms into the zeolite framework with heavy adsorbing compound molecules. Here, one can use the General Utility Lattice Program[3] to perform convenient grand canonical Monte Carlo (MC) simulations. Then molecular dynamic (MD) simulations can be used for reproducing of diffusion of helium atoms in differently loaded with heavy adsorbing compound zeolite framework . This can be done using DL_POLY code[4]. Analyzing accumulated statistical data by calculation of mean square displacement (MSD) of given species one can to calculate the self-diffusivities and then to approximate the permeance of helium and heavy adsorbing compound molecules. And the final step is to visualize the computational experiment results. Realizing such complex process is a time consuming, demanding a lot of human-operators' attention, from job submissions till processing the results of particular running, transforming data and generation of input files. Automation of this work can give a computational science plenty of benefits.

[*] Partially supported by the RFBR grant 10-07-06800-mob_g

The concept of Grid computing is one of the possible solutions to organize such collaboration of scientific applications. The conceptual basis of Grid technology is to combine resources by creating a new type of computing infrastructure, providing global integration of information and computing resources based on network technology and special software intermediate level (middleware) as well as a set of standardized services to ensure a reliable shared access to geographically distributed information and computing resources: individual computers, clusters, storage and information networks.

At the moment, there are number attempt to creation the framework for aggregating of various computational science applications and automation of creation of the workflows. For example, Accelrys Material Studio an MyGrid project. These projects provides a number of tools and workbench for the supporting of computation running, processing and deployment of computation results.

In the present paper we discuss how the concept of grid computing can be used in deploying of complex workflows. We consider various strategies to development and implementation of the framework and workbench for the supporting of complex simulation runnings in computational science areas. This computational suite should aggregate different simulation and computation software to be able performing a variety of tasks in computational physic, chemistry and material science area, based on various methods, such as force field, density functional theory, continuous and percolation models, etc. This suite should allow to automate running of the multistage calculations using an existing computation science software, with arbitrary computation chains of applications – workflows, and possibility of its algorithmization.

## 2  The Problem Statement

Proposed system should include up-to-date technologies and concepts, and satisfy following important principles:

**Interoperability of computational software.** The main goal of developing of the new computational suite is to provide an opportunity to utilize existing computational software, as widely used like GULP, DL_POLY, GROMACS, SIESTA, NWChem, etc., so some handcraft utilities and tools. This raises the problems how to provide essential data from the results of one applications onto the input of another one, how to store, analyze and visualize intermediate and final results. This question leads to developing the general- purpose representation of essential sets of parameters describing arbitrary model of the simulated system. Here, "essential" means a such set that contains full information about the system required to start any computational application. This is a set of instant and local observables.

**Distributed informational system.** It is necessary to ensure the coordinated use of computational resources, databases and repositories in the absence of centralized control of these resources. It should be taken into account the possibility of freely adding new resources and scalability of the system, respectively, as well as, possibility of the use of existing resources (eg, crystallographic databases)

**Local autonomy.** A particular resource - the components of the system, as its fragment, is also a complete local resource, perhaps with its administration policy.

**Independence of nodes.** System nodes are equal in obtaining systems services and independent of each other. At any time, the node can withdraw from the scope - a dynamic property of the environment. With this in mind, you want to ensure achievement of assigned tasks and the ability to recover from errors associated with the node withdrawal.

**Contiguous operation.** It is necessary to provide opportunity for continuous access to resources regardless of their location and regardless of the operations performed on local sites.

**Transparent location.** The user is abstracted from the physical location of data, the specific properties of resources and hardware features of computing system. System nodes are diverse having different hardware, operating systems and databases. Data models, formats, databases, batch systems and system queues, can vary from resource to resource.

**Transparent network.** Access to any resource over the network provides using standard, open, general-purpose protocols and interfaces.

**Quality of service.** It is required to provide specified level quality of service, using the system resources so that it can be possible to maintain a certain quality of service on various parameters. Then, integration of resources of various types, ready to fit the complex needs of users, will benefit the use of this combined system is much higher than the sum of its parts.

**Collective intelligence.** It should be able to share obtained results, algorithms and techniques of computational experiments and treatments between users of the system. This possibility should stimulate the accumulation, enhancement and publication of knowledge. The suite should also provide the possibility of maintainance the collaboration of many participants in carriing the computational experiments out and a group development of software and computational workflows.

**Reproducibility in computational science.** Currently, the issue of reproducibility in computational science is becoming more and more acute, due to the fact, that modern science increasingly depends on the numerical experiments (simulations), the numerical models and computational analysis of large amounts of data. Computational methods are reproducible in principle, but calculations, that is reproducible in principle, cease to be reproducible in practice with the increasing complexity of modeled systems and the increasing amounts of data. A third party independent scientist, who refer skeptically to the results of particular calculations, for its verification, should spend considerable time to install the necessary software, debug workflow of the computations, and perhaps, find particular parameters of the calculations that are hidden for reason of their seeming insignificance, because of the limited volumes of the publication or any other causes. The opportunity to publish the workflows and the availability of a system capable to reproduce it, as well as possibility to store options of computation, will partially solve the problem of reproducibility of the numerical experiment.

**Security.** Besides the usual issues of information security, there are a few more. Computing software is usually not free. Even if it is not commercial, most often it is distributed under the academic license. Any license imposes some restrictions on the use of software distributed with it. Thus, during the formation of workflows it should be considered, as what license agreements related to the user, so what user can do regarding these agreements.

Another issue is formation a list of citation. When a computational experiment is performing, parameters of this experiment, its methodology and procedure(workflow), mostly, has been taken from someone else's previous work, with, perhaps, minor modifications. Therefore, it is necessary to keep records of what work, whose and where, was used in the experiment. It is also necessary to develop a mechanism that prevents the possibility of refusal on the fact of using someone else's intellectual work in their work.

**Cloud and pervasive computing.** The device(PC, mobile device, the browser) using by the user to work with the system is a cache, that reflects the current work. It is like to having an analog of a full-featured desktop that is accessible from any node connected to the Internet. And like a thin client for remote access to the system through the net.

## 3 Integration of Computational Software

At this time there are a large amount of various computational software of any kind. Porting of any one of them to the grid platform making it grid-enable is very consuming. Other possible way to introduce existing software to the grid is to wrap it in a grid service. This approach seems to be more convenient. Encapsulation of the convenient application as a service provide the basis for achieving interoperability between different software and allows to building and execution of any workflows.
To be able to provide data required for starting new application onto its input from the output of previous application, we need to extract essential information from the files with the results. This information is a set of various physical values describing the properties of modeled system. The units of these values varies from application to application. So, we need to convert them, and then, we can form the input files for a next in workflow application. There are as many formats of these input and output files, as applications. A very few application can read foreign data files. As a result, we need to provide intermediate components to transfer data from one application to another, that would convert the presentation of data. This approach has not good scalability. Another way is to use some universal representation of physical values. In this case, data needed to be transmitted, at first, should be converted to the intermediate format, and then can be converted from intermediary to the required one. Processing workflow, the system can store the results in that intermediate format after each application completion. This feature gives also such advantage as possibility to providing the rollback functionality and additional reliability of the system.

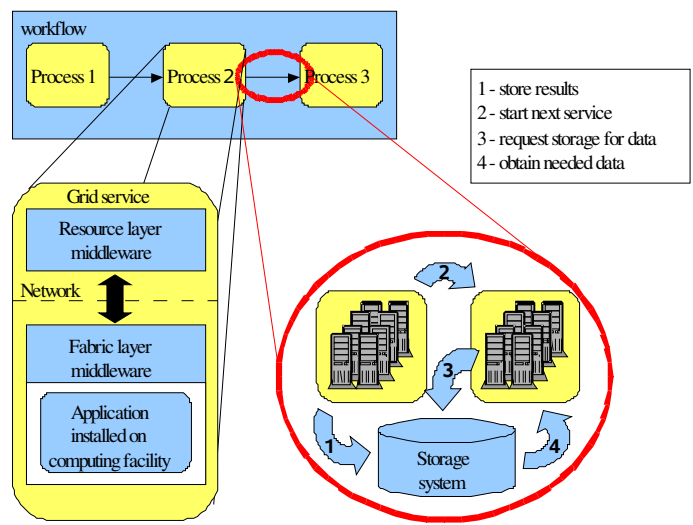

**Fig. 1.** Implementation of workflow: computing grid resource architecture and services interaction concept.

Concerning data transmitting in accordance with a particular workflow in computational science, full amount of data, that can be available from the output of previous application, is often redundant. Therefore, each succeeding process in workflow would extract essential for it data from stored by predecessors results.

There is also possibility to make decision what kind of data should be stored governed by succeeding succeeding processes in workflow. Let us consider these tasks in terms of the open architecture of Grid services (OGSA)[5], when the system is based on the interaction of a set of independent or loosely coupled services. Computer software, the services of which we need for each step of the workflow, should be regarded as a grid service. This service will be a generalization(virtualization) of such resource as a particular software. In order to deviate from the questions of specific installation of particular software on the computing resources (PC, cluster, grid infrastructure), the computing facility(calculator) should also be included in this service. Thus, a particular service that is the representation of such a resource as a particular calculator + stand-alone program in a generalized form. This representation allows us to abstract from such inessential differences in resources as a way to start the program (job submission), placing the input and output files, etc. Returning to the five-layers view of Grid protocols architecture[6], this resource is at the fabric layer, and to be able to deliver its services to others, should have its own representation at the layer of resources. Thus, it should be developed middleware of fabric layer being the object, which encapsulates a complex resource of such kind as calculator + program. On fig.1 the sketch of such approach is shown. Considering that the functionality of this middleware is to obtain a job, run the program, pending the completion, transfer of results, etc., are not principal to the differences in resources (the difference between programs and the difference between calculators), it should be abstracted from it. It is necessary to separate this core functionality from the immediate action to launch a separate program on a separate computer. We should formalize these immediate actions, through the use of templates, and language transformation, such as using XML for template and XSLT to convert it to a particular platform.

We should also develop middleware for representation of these fabric-level objects at the layer of resources. It should allow to virtualize resources, providing the ways to control over them and its detection. Here is takes place the initiation of the resource, monitoring and accounting of its use, mutual approval of security policies of resource utilization. Service is being built on the basis of such resources layers' middleware should has a tight semantic and strictly specified interface. As well as at the fabric layer, at resources' layer, the main functionality can be separated from functionality, specific to an individual resource, specifying the interaction with individual fabric component uniformly for all resources. Interaction of fabric and resource layer middleware is carried out via standard and open Internet protocols. Thus, every particular program installed on a computer system (computer, high-performance system) is a standalone resource (calculator + program) which will be presented by a pair of fabric and resource layer processes interacting through a network(see fig.1). Both processes are implemented in two middleware applications common to all the resources. And the representation of this component at the resource layer is a grid service providing a standardized service.

To ensure the interoperability of computer software, it is necessary to organize the interaction of these services. As noted earlier, to transfer the results of one program to the input of another, one should perform the transformation of data formats and files.

As it has been showed early, we need an intermediate format which satisfies to all the formats of software included in the system. This allows to perform a multi-step tasks by implementing the transformation of data from the particular program data format to the internal and, conversely. Converting to and from the intermediate format can also be formalized, thus increasing the versatility of the system. The services interaction should be arranged in the same way - through an intermediary. In this case, the mediator will be a storage system. It should be developed a middleware that is the interface to the storage which will store the results of programs in an intermediate format. This service should have the same interface as computing services. Thus, interoperability of computing services will be provided in followed way. One computing service save its work in common for interacting services storage subsystem and then, another one would take the necessary to it data. Such an approach would not only solve the problem of interoperability of computing software, but will, due to the already existing results in Grid technology, to solve the problems of distributed information system creation, local autonomy, the independence of nodes, contiguous operation and transparent location.

## 4  Graphical Representation of Workflows

For improving quality and speed of design of computational experiments workflows we suggest to include instruments of graphical description of workflows into the developing system.
Activity Diagrams (AD) can be used as a basis of graphical language for this purpose. AD are graphical representations of workflows of stepwise activities and actions with support for choice, iteration and concurrency. AD can be used to describe step-by-step workflows of components in a system. AD representation is intuitive for a designer. Besides, a lot of parallel flow graphs analysis methods, developed in last decades, can be applied to AD representation of workflow. Particularly, additional features for verification of workflows in logical level appear [7].
Lets view an AD in Unified Modeling Language (UML AD) [8]. Rounded rectangles (activities), diamonds (decisions), bars (start or end of concurrent activities), black circle and encircled black circle (the start and the end of the workflow) and arrows (control flows) are the constructive components of UML AD. Data transferring between the activities are labeled in the diagram with the object nodes (rectangle, attached to the oriented edge and inclusive of the name of the object) or with fixing of input and output (the names of incoming and outcoming objects are labeled at the activities).
To denote the distribution of roles the area, occupied by UML AD, is divided into swimlines. Only those activities, for which the concrete object is responsible for, are displayed on the swimline. In this case the services should be considered as the objects. But the tasks, for which the developing system is aimed at, are of such kind that the number of the actuators and number of the works are the same, so only one activity will be placed at each of swimlines while standard usage of swimlines. A large number of swimlines in the diagram disimproves its clearness (it is important while design, presentation of researches results to scientific community and so on). Besides, if the system is well-developed, a designer does not have to define the actuators – the system should find necessary services itself. On the other side, swimlines can be used in other way than in UML. For example, only those works can be placed on the first swimline for which a designer defines either a program and a actuator; on the second swimline only those works for which a designer defines only a program but an actuator is chosen by the system; on the third swimline only those works for which the search is fulfilled by the system. So the visual design tool should provide an ability to a designer to choose AD notation. Such variants are possible: with the representation of swimlines as in UML; with more compact representation of the services (directly in the activities, for example); without defining of services and some others.

AD for modelling of diffusing of helium incorporated into a mixture of strongly adsorbing components (hexane, benzol) through the pores of zeolite membrane is shown on fig. 2. Crystallographic data (CD) – coordinates of atoms in crystal – are extracted from the zeolite structures database [1], and the force-field potentials of atom and molecular interaction are taken from publishing or comparison with the experimental data. It is supposed, that the stirring of a strongly adsorbing component is insignificant in the comparison to light helium, that is why the task of adsorption of a heavy component should be solved primarily. The arrangement of its molecules in zeolite lattice after adsorption is modeled by the CBMC method . For operating simulations the program pack CBMC of calculations BIGMAC [2] is used. The result of the calculation is the system configuration – coordinates of atoms of zeolite and adsorbate. Further the distribution of atoms of helium surrounded by zeolite lattice and molecules of adsorbate. That is the task of adsorption as well and it is solved by the MC method either in the system of grand canonical ensemble. However recently the General Utility Lattice Program (GULP) is used for the simulation [3]. CD of zeolite, adsorbed component and helium are the result of this usage. For the diffusion modelling classical MD method is applied, in which atomic motion is found by classical Newton equation. The simulation of MD is fulfilled with DL_POLY program [4]. During simulation statistic information about the system is stored and self-diffusion coefficient is calculated by the computation of the mean square displacement of atoms of each species.

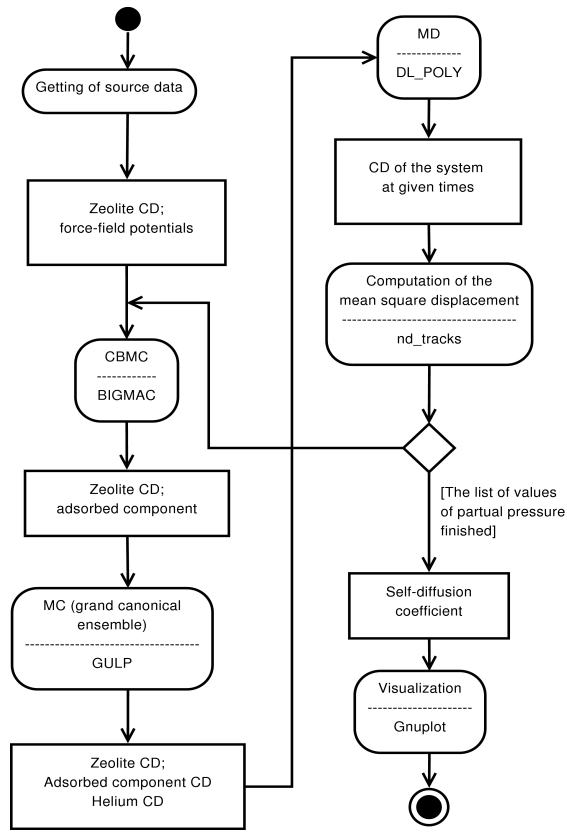

**Fig. 2.** AD for modelling of diffusing of helium incorporated into a mixture of strongly adsorbing component through the pores of zeolite membrane.

The concept of of all-round computations is one of the basic in project. Particularly it is realized by by the rendering of full-function web-interface to the system. So workflow visual design tool should be constructed as a web-application, i.e. client-server application, and its functions are arranged between a server (web-server) and a client (web-browser of designer) and the data storage is mainly server-based. Client part can be realized on JavaScript or with the applying of some applets (Java, Flash and others).

Translation AD to xml-description of sequence of the activations of works is a compulsory functional requirement to workflow visual design tool. The Job Submission Description Language (JSDL) is being used for description the requirements of individual computational jobs for submission to resources, particularly in Grid environments. JSDL does not describe the relationship between individual works described using JSDL, and in turn, the relationship of those works with the data they consume and produce. Consider the negotiation required to put a job in an optimal resource environment according to its requirements is as well beyond the framework of JSDL. In this case, the use of appropriate negotiation and agreement protocols are required [9].

The formal representation of workflow is offered to fulfill as a program in functional programming language Haskell [10]. Besides compactness and formal preciseness of description one of the advantage of this approach is the capability of the usage of in-built mechanisms of parallelization into Haskell (map and reduce). Thus the task of AD translation into a program in Haskell and vice versa appears, moreover the task is purely technical – the rules of translation of parallel graph chart to a program in a functional programming language are well-known, see for example[11].

## 5   Conclusion

In this paper we study some aspects of the implementation of grid-based framework for establishment of computational workflows. The applications comprising workflow are not need to be grid-enable. It has been described approaches to encapsulate convenient application into grid service, and strategies for enabling interoperability between them. We highlight the main requirements addressed the developing of such framework. The UML based instruments of graphical description of workflows for the developing system has been suggested. As case study, creation of workflow for the modeling of helium diffusion through a zeolite pores loaded with heavy adsorbed compound has been examined.

11.Kutepov, V.P., Fal'k, V.N., Forms, Languages and Complexity Parameters of the Parallelism. Software and Systems(in Russian), №3 (2010)